\documentclass[aps,prl,twocolumn,showpacs]{revtex4}

\usepackage{exscale}
\usepackage{graphicx}
\usepackage{amsmath}
\usepackage{latexsym}
\usepackage{amsfonts}

\DeclareMathOperator{\Tr}{Tr}
\DeclareMathOperator{\conv}{conv}
\DeclareMathOperator{\cl}{cl}
\DeclareMathOperator{\sign}{sign}
\newcommand{\ket}[1]{ |  #1  \rangle}
\newcommand{\bra}[1]{ \langle #1   |}

\DeclareMathOperator{\eins}{ {\bf 1}}
\begin{document} 

\title{
A two-way algorithm for the entanglement problem
}

\author{Florian Hulpke and Dagmar Bru\ss }
\affiliation{
Institut f\"ur Theoretische
Physik,  Universit\"at Hannover, D-30167 Hannover, Germany}

\date{Received \today}

\begin{abstract}
We propose an algorithm which proves a given 
bipartite quantum state to be separable
in a finite number of steps. Our approach is based on the 
search for a decomposition via a countable subset of 
product states, which is dense within all product states.
Performing our algorithm simultaneously with the algorithm
by Doherty,  Parrilo and Spedalieri 
(which proves a quantum state to be entangled in a finite number of steps) 
leads to a two-way
algorithm that terminates for {\em any} input state.
Only for
a set of arbitrary small measure near the border
between separable and entangled states the result is inconclusive.
\end{abstract}
\maketitle

The question of whether a given quantum state is entangled or
separable is  both of fundamental interest, and of 
 relevance for  the implementation of
quantum information processing tasks.
The separability problem 
has stimulated many ideas for partial solutions:
A sufficient condition for separability is given
by the  vicinity of the state to the identity
\cite{volume1,volume2,volume3}. A necessary condition for
separability of a given state is that it fulfills the
criterion of the positive partial transpose (PPT) \cite{PPT}. 
Entanglement witnesses provide 
sufficient criteria for entanglement
\cite{witness}. 
 However, the separability problem
 has been shown in 
\cite{NP} to be in the complexity class NP-hard,
 and no complete solution is known yet.
An improved algorithm for the separability problem,
based on entanglement witnesses, was recently proposed in
 \cite{artur}.
In this Letter we  suggest an algorithm that
extends and complements 
the recent algorithm by
Doherty, Parrilo and Spedalieri 
\cite{Parrilo1}.

The separability problem is  defined as follows. 
A quantum state $\rho$ which acts on a bipartite, finite-dimensional 
Hilbert space ${\cal H}_A \otimes {\cal H}_B$ 
is {\em separable} iff there 
exists a set of 
pure product states $|e_i\rangle\langle e_i| \otimes 
|f_i\rangle\langle f_i|$,
and a set of real positive numbers $p_i$
with $\sum_i p_i=1$, 
such that $\rho= \sum_{i} p_i |e_i\rangle 
\langle e_i | \otimes |f_i\rangle \langle f_i|$ \cite{Werner}. 
This property can be reformulated such that
 $\rho$ has to lie within the convex hull of some
pure product states.
Furthermore, it is  known that a separable 
$\rho$ is  in the
convex hull of at most $L:=(\dim{\cal H}_A \dim{\cal H}_B)^2$ pure 
product states \cite{cara-horo}. 
So it remains to show whether for a given $\rho$ 
there exist $L$ (not necessarily pairwise different) pure product states, such
that $\rho$ is in their convex hull.
However,  searching for these $L$ states 
in the set of all quantum states 
would mean to search through a set of infinitely many 
{\em uncountable} states.

One of the most advanced solutions to the separability
problem was recently introduced  by 
Doherty, Parrilo and Spedalieri 
in \cite{Parrilo1}. They presented an iterative algorithm 
(denoted as ${\cal A}_1$ in
the remainder of this text), which is  based on 
symmetric extensions of a 
given quantum state, such that this algorithm  terminates
after a finite number of iterations 
iff the state is {\em entangled}.
However, if the state is separable, the algorithm ${\cal A}_1$ does not
terminate.
From an algorithm-theoretic point of view this is not 
satisfactory: not having terminated after a finite time does not
yield any information about the properties of the state.

We suggest an algorithm which provides a solution for this problem
and closes the gap in the above algorithm,
because it detects a given {\em separable} state after a finite number of
steps. 
 Applied in parallel with the algorithm in \cite{Parrilo1},
the combined  algorithm then
 terminates after a finite time: one of the two tests 
certainly
terminates, as every state is either entangled or separable.
Our main idea is that it is sufficient to restrict ourselves to a 
{\em countable} subset of pure states, rather than searching through
all (uncountable) pure states.

Let us start by providing the mathematical background.
For each  
separable state $\rho$
 there exists by definition
 a  decomposition
$\rho= \sum_{i=1}^L p_i |e_i\rangle 
\langle e_i | \otimes |f_i\rangle \langle f_i|$, 
with $p_i\geq 0$ and $\sum_ip_i=1$.
A central idea of our approach is  that the probabilities $p_i$ 
are not needed for deciding whether a state is separable or
not -- only the pure states $\{\ket{e_i},\ket{f_i}\}$
in the decomposition are
essential. Thus we can rephrase
the separability definition in the following way:
 {\em A state $\rho$ is
separable, iff there exists a set of 
projectors onto separable pure
states $ c:=\{|e_1\rangle\langle e_1| \otimes |f_1\rangle \langle
f_1|,...,|e_L\rangle\langle e_L| \otimes |f_L\rangle\langle f_L|\}$,
such that $\rho$ lies in  the convex hull of the elements of
$c$}.
However, a straightforward search 
through all sets containing $L$ pure product states
- while easily parametrized - is impossible, as there are
  not only infinitely many such sets, but they are even uncountable.

We will show in the following that it is  (in the generic case) enough
to restrict oneself to a countable subset $C$ of all $L$-tuples of
pure product states, which is  
dense within all pure product states.  
Within a countable set $C$ there exists
by definition  a sequence $\{c_i\}$  in $C$ that
covers  $C$ completely. Thus
 it is  possible to formulate an iterative algorithm that 
passes {\em all}
$L$-tuples of pure product states in $C$ 
in the limit of infinitely many steps. 
 Furthermore 
one can use  the well-known 
feature  
that for every element $c$ in $C$ 
there exists 
a finite number $i$ such that 
the value $c_i$ at the $i$-th step of the sequence
 equals $c$.
Therefore for a given (generic) separable $\rho$ 
the iterative algorithm: {\it 
Check whether $\rho$ is in the convex hull of the $L$-tuple
$c_i$}  terminates after a finite 
time. 

Let us put the mentioned ideas on  mathematical grounds and
first
prove the statement that it is sufficient to search within  a dense
subset. Although some of the arguments given below hold for
general convex sets, we will restrict ourselves 
here and in the following to sets of operators
(``states'') acting on a finite-dimensional Hilbert space, where 
we use 
the  
Hilbert-Schmidt norm. The  distance  between two 
vectors from the Hilbert space is given by
$d(|\psi\rangle, |\phi\rangle) =\sqrt{(\langle \psi| -\langle
\phi|)(|\psi\rangle - |\phi\rangle)}$, and between two operators
$d(a,b)= \sqrt{\Tr[(a^\dagger-b^\dagger)(a-b)]}$.

{\bf Def. 1:}  
A subset $B \subset A$ is called {\it dense} in $A$ if 
every $a \in A$ can be written as the limes of a sequence $\{b_n\}$ in $B$. 

Let us  summarize some facts about {\em convex} sets: \newline
1. A set $X$ is convex if for any finite number $l$, any  $a_1,...,a_l \in X$, 
and all $\lambda_i
\geq 0$ with $\sum_i \lambda_i=1$, the sum $\sum_{i=1}^l \lambda_i a_i
\in X$. \newline
2. The convex hull of a set $X$ is the smallest convex set that contains
$X$. The convex hull of the set $X$ will be denoted as $\conv X$. \newline
3. For a convex set $X$ we will denote the border $\delta X$ 
as all points $a \in X$ for which 
$ \exists \, b \in X$ such that for all 
$\eta>0$ the point $(1+\eta) a - \eta b$ does not belong to $X$.

{\bf Lemma 1:} 
Let $A= \cl A$ 
(where $\cl A$ denotes the closure of $A$)
be the set of extremal points of a convex set $X$, and 
$B \subset A$ be a dense subset of these extremal points, then
the convex hull of $B$ is dense within the convex hull of $A$.

{\bf Proof:}
We want to prove that $\conv A$= $ \cl \conv B$,
so we have to show both inclusions.\\
``$\subset$":
Suppose one has  $x \in \conv A$ then there exists by definition
a set of elements $a_i$ in $A$ and $\lambda_i>0$ with $\sum_i
\lambda_i=1$,
such that 
$x=\sum_{i=1}^r \lambda_i  a_i$. 
Since $B$ is dense in $A$ there exists for every $a_i$ a sequence $\{b_{i,j}\}$
in $B$ such 
that the limit  $j \mapsto \infty$ of this sequence is $a_i$.
Due to the additivity of the limes 
$x$ is  in the 
closure of $\conv B$.\\
``$\supset$": If $A$ is closed then $\conv A$ is also closed, and contains 
$\conv B$. But $\cl \conv B$ is the smallest 
closed set that contains $\conv B$, so
$\cl \conv B \subset \conv A$. \hfill $\square$

{\bf Lemma 2: } 
Given a convex set $X$ within the set of states, with the finite set
of extremal points $a_1,a_2,...a_l$  
and a point $x$ that is in the interior of $X$. Then there is an 
$\varepsilon>0$
such that for all points $a'$ with $d(a_1,a')<\varepsilon$  
the point $x$ is also in the interior of the convex hull of the points
$a',a_2,...a_l$.

\begin{figure}
\begin{minipage}[c]{.45\textwidth}
\centerline{\includegraphics[width=.6\textwidth]{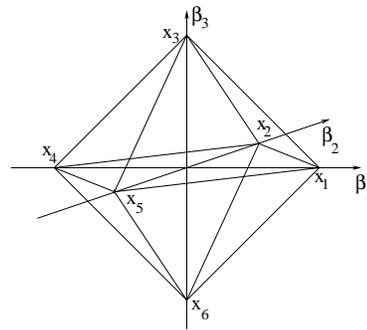}}
\vspace*{13pt}
\caption{\label{fig2}Constructing a ``generalized diamond'', 
i.e. a regular crosspolytope.}
\end{minipage}
\end{figure}

{\bf Proof:}
A ``face'' of a convex set with
dimension $D$
is a subset of the border, that is the convex hull of $D$ affinely
independent elements. 
Since $x$ is in the interior of $\conv \{a_1,..a_l\}$, there is an 
$\varepsilon>0$ such that all points $y$ in $X$ 
for which $d(x,y)<\varepsilon$
are not on the border of $X$. 
In other words:
 $x$ is not in an $\varepsilon$-surrounding of the border.
We note now that  changing $a_1$ to  $a'$  only affects those
faces that have $a_1$ as one of the affinely independent extremal points.
When $a'$ is taken such that
$d(a_1,a')<\varepsilon$, the involved faces will remain in an 
$\varepsilon$-surrounding 
of  the faces of the original set of extremal points. 
Therefore
 the  point $x$  remains in the interior of the new
convex set.
\hfill
$\square$
\newline
{\bf Theorem 1:} 
Given two convex sets $B \subset A$, such that $B$ is
dense within $A$, then 
all points in the interior 
of $A$ are in  the interior of $B$. \\
{\bf Proof:}
Consider an arbitrary point $a$ 
in the interior of $A$. 
Then 
$\exists\, \epsilon >0$ s.t. all points 
with a distance smaller than $2 \epsilon$ from $a$ are in $A$. 
Let us choose $2l$ points defined by $x_i= a+\epsilon \beta_i$, 
$x_{l+i}= a-\epsilon \beta_i$, with $i=1,...,l$, where
$\{\beta_1,...,
\beta_l\}$ is an orthonormal basis
in our space. Each of these 
points has the distance 
$\epsilon$  
from $a$, and is 
therefore 
in  $A$.
 The convex hull of these $2l$ extremal points is
a ``generalized diamond'', i.e. a regular crosspolytope
 (see Fig. \ref{fig2}),
and $a$ has the distance $\sqrt{l}\epsilon>0$ from 
any face of this diamond. 
Using Lemma 2, we can shift one extremal point 
of the diamond after the other,
such that the shifted points are in $B$
(as $B$ is dense in $A$),
while keeping $a$ in the interior of the new diamond.
As 
 all new extremal points  belong to 
$B$, $a$ is in the interior of $B$.
\hfill
$\square$

Let us  apply these general properties of convex sets to
the problem of proving the separability 
of a given $\rho$. We will  parametrize a 
countable subset of pure product vectors that is dense within all
 pure product vectors,
and show that this leads to a dense subset of corresponding
1-dimensional projectors. 
 Here, we have to distinguish the cases
where $\rho$ has full rank, or does not have full rank.

We first study the case that $\rho$ has  full rank.
The set of all pure product vectors is parameterized by  fixing
an orthonormal
basis $\{|1\rangle_A,...,|n\rangle_A\}$
in ${\cal H}_A$, where $\dim {\cal H}_A=n$, and an orthonormal basis  
$\{|1\rangle_B,...,|m\rangle_B\}$
in ${\cal H}_B$, where  
$\dim {\cal H}_B=m$.
Then every pure product state can be written as
$
|\psi\rangle = |a\rangle \otimes |b\rangle = 
(\sum_i \lambda_i |i\rangle_A) \otimes (\sum_j \mu_j
|j\rangle_B), 
$
where $\sum_i |\lambda_i|^2=1$ and $\sum_j |\mu_j|^2=1$. So the pure
separable states are parametrized by the set $G=
\{\lambda_1,...,\lambda_n,\mu_1,...,\mu_m | \sum_i |\lambda_i|^2=1,
 \sum_j |\mu_j|^2=1
\}$
of $n+m$ complex coefficients with the two normalization constraints.

We  now restrict the coefficients to 
 complex quantities expressed with
 rational numbers $\mathbb Q$.
Note that 
the rational numbers  are a countable set and dense in the
set of real numbers $\mathbb R$. However,
due to the normalization constraint we cannot  simply consider
  the subset of $G$ where all coefficients are of the form 
$\frac{p}{q} + {\bf i} \frac{r}{s}$ (where $p,q,r$ and $s$ are natural numbers).
We  solve this problem by
embedding the normalization constraints explicitly,
choosing  the subset 
\begin{equation*}
{\tilde G}=\left\{
\left.
\begin{tabular}{l}
$\lambda_j=\frac{p_j}{q_j} {\bf e}^{2 \pi {\bf i} \frac{r_j}{s_j}}$ \\
$\lambda_n=\sqrt{1-\sum \limits_{l=1}^{n-1} \frac{p_l^2}{q_l^2}}
{\bf e}^{2 \pi {\bf i} \frac{r_n}{s_n}}$  \\
$\mu_{k}=\frac{p_{n+k}}{q_{n+k}} {\bf e}^{2 \pi {\bf i}
  \frac{r_{n+k}}{s_{n+k}}}$ \\
$\mu_{m}=\sqrt{1-\sum \limits_{l=1}^{m-1} \frac{p_{n+l}^2}{q_{n+l}^2}}
{\bf e}^{2 \pi {\bf i} \frac{r_{n+m}}{s_{n+m}}}$
\end{tabular}
\right|
\begin{tabular}{c}
$1\leq j \leq n-1$ \\
~\\
$p_i, q_i, r_i,s_i \in {\mathbb N}_0$ \\
$p_i\leq q_i, r_i \leq s_i$\\
$1\leq k \leq m-1$ \\
~\\
~\\
\end{tabular}
\right\}.
\end{equation*}
The subset $\tilde G$ is dense within $G$, since for every element
  $g=(\lambda_1',...,\mu_m')$ in $G$ there is an element  ${
  \tilde g}=(\lambda_1,...,\mu_m)$ in $
\tilde G$ that is 
arbitrary close to $g$, when  the distance is defined as
$d( g, {\tilde g})=
  \sqrt{(\lambda_1'-\lambda_1)^2+...+(\mu_m'-\mu_m)^2}$. 
Furthermore $ \tilde G$ is
 countable, since it is a subset of
${\mathbb Q}^{\times 2(n+m-1)}$.

Obviously the product vectors 
parametrized by 
$\tilde G$ are dense within all product vectors in  the Hilbert space,
since
the distance $d( g, {\tilde g})$ 
is equal to the distance induced by the Hilbert-Schmidt norm.

{\bf Lemma 3:}
If a sequence  of normalized 
vectors $|\psi_i\rangle$
converges towards  $|\phi\rangle$, then the corresponding projectors
$|\psi_i\rangle \langle \psi_i|$ converge towards the projector
$|\phi\rangle \langle \phi|$.

{\bf Proof:} 
The distance between $|\psi_i\rangle$ and $|\phi\rangle$ is 
$d(|\psi_i\rangle, |\phi\rangle)
=\sqrt{2( 1- {\Re  [\langle \psi_i|\phi\rangle]})}$,
where we denote by $\Re[\langle \psi_i|\phi\rangle]$ 
the real part of the scalar product.
Since the $|\psi_i\rangle$ are converging towards $|\phi\rangle$,
for every $\epsilon >0$ there exists an $i_0$ such that for all
$i>i_0$ the distance $d<\epsilon$. This implies that 
$\Re [\langle \psi_i| \phi\rangle] > 1- \epsilon^2/2$.
The distance 
of
the corresponding operators is calculated as
\begin{align}
d(|\psi_i\rangle \langle \psi_i|, |\phi\rangle \langle \phi|)
& 
= \sqrt{2(1- |\langle \psi_i|\phi\rangle|^2)} \\
\nonumber
& \leq \sqrt{2(1- (\Re [\langle \psi_i| \phi\rangle])^2)}
\leq\epsilon\sqrt{ 2 - \epsilon/2}\ .
\end{align}
Thus 
 $|\psi_i\rangle \langle \psi_i|$ converges towards $|\phi\rangle
\langle \phi|$.
\hfill 
$\square$

We now study the case that $\rho$ does not have full rank.
Since the states with lower
rank form the border of all states,
they
will not necessarily
be in the convex hull of the
previously defined countable set. 
Thus we have to define
the set 
$\tilde G$ in a different
way.

Let $r$ be the rank of $\rho$. 
We restrict ourselves to 
the $(r^2-1)$-dimensional space of 
Hermitean operators with trace 1 
that are supported at most on the range of $\rho$.
Therefore the maximal number
of extremal points needed to find a separable decomposition 
is given by $L:=r^2$. 
We know that $\rho$ is in the interior of the space
spanned by the projectors whose corresponding vectors are in the
range of $\rho$.
Thus, in the case of less than maximal rank we
do not have to check whether
$\rho$ is in the convex hull of  all separable pure states, 
but   whether $\rho$ is in the convex hull of all 
product projectors, whose  vectors are in its range. 
Note that the existence of ``enough" product vectors in the range 
 is a necessary, but not sufficient criterion for separability, since
any state of full rank has all pure states in its range,  
independently of its separability property.

Given a state $\rho$ 
and its spectral decomposition 
$\rho=\sum_{i=1}^r p_i |\phi_i\rangle \langle \phi_i|$, a
 vector  $|\psi\rangle$ in the range of $\rho$  can be written as
$|\psi\rangle= \sum_{i=1}^r \lambda_i |\phi_i\rangle$, with
complex coefficients $\lambda_i$ and 
$\sum_{i=1}^r |\lambda_i|^2=1$.
A pure bi-partite state $|\psi\rangle$ is separable iff
$Tr_A (Tr_B |\psi\rangle\langle\psi|)^2=1$. 
Therefore the coefficients for all pure 
product states in the range of $\rho$
are the roots of a polynom of fourth order.
These roots can be obtained numerically.
The conditions for  the $\lambda_i$ are summarised as follows:\newline
1.
$\lambda_i \in {\mathbb C}$ $\Leftrightarrow$ $\lambda_i = {\bf e}^{
2 \pi {\bf i} \theta_i} |\lambda_i|$ \newline
2.
$\sum_{i=1}^r |\lambda_i|^2=1$ $\Leftrightarrow$ $|\lambda_r|=\sqrt{1- 
\sum_{i=1}^{r-1} |\lambda_i|^2}$
\newline
3. 
$\Tr_A (\Tr_B |\psi\rangle \langle \psi|)^2 -1=0$, i.e.
\begin{eqnarray}
&&\sum_{j,j'=1}^n\sum_{i,i'=1}^m\sum_{l,k=1}^{r}\sum_{l',k'=1}^{r} 
\lambda_l \lambda_k^* \lambda_{l'} \lambda_{k'}^*
\bra{j_A i_B}
 \phi_l\rangle \langle \phi_k\ket{j'_A i_B}\cdot \nonumber \\
&& \hspace*{1.6cm}\cdot \bra{j'_Ai'_B} \phi_{l'}\rangle \langle \phi_{k'}
\ket{j_Ai'_B} -1=0
\end{eqnarray}

We  parametrize the product vectors 
in the range of $\rho$ by $2r-2$ real
parameters.
Now we once again construct a dense subset of these parameters
by choosing rationals in the form $\lambda_i = \frac{p_i}{q_i} {\bf e}^{2\pi
  {\bf i} \frac{r_i}{q_i}}$, where $p_i,q_i,r_i,s_i \in {\mathbb N}$ and
$p_i <q_i, r_i<s_i$. As previously, this subset is  dense in all product
vectors and therefore dense in the corresponding projectors. 
Furthermore it is countable, due to the countability of the rationals.

Having found a countable subset that is dense in all product states,
this immediately leads to 
a countable subset $C$ of $L$-tuples of product states:
each fixed $L$-tuple $c_i$ can be provided with a ``finite address''
$i$.
 We can restrict ourselves to tuples with
affinely independent elements: 
An affinely dependent set can be reduced to an affinely independent subset.
For a decomposition with less than $L$ product states,
one can extend the corresponding tuple to an $L$-tuple by
adding affinely independent entries. 
Obviously the state is still in the convex hull of the extended
tuple, and will be detected as separable at its ``address''.

Our arguments lead to an 
 algorithm
for the detection of a separable state, 
 in the following denoted 
as ${\cal A}_2$:
One walks step by step through the countable set of $L$-tuples $C$.
The $i$th-element of $C$ is
${c_i}=\{\tau_{1}^{(i)},...,\tau_{L}^{(i)}\}$.
One checks if the 
$\{\tau_{1}^{(i)},...,\tau_{L}^{(i)}\}$
are affinely independent, and if they are not, one moves to the
next element, i.e.
$i \mapsto i+1$. 
If the elements of $c_i$ are independent, one checks whether
$\rho$  belongs to $\conv \{\tau_{1}^{(i)},...,\tau_{L}^{(i)}\}$. 

The check whether $\rho$ is in $\conv \{\tau_{1}^{(i)},...,\tau_{L}^{(i)}\}$
is performed as follows:
one chooses  $L-1$ different elements out of the
$L$ given ones,
and finds the normal $\xi_n$ to the hyperplane defined by these
elements.
The state $\rho$ is on the ``same side'' of the hyperplane as
the remaining point,
if $\sign(Tr[\xi_n\rho])=
\sign(Tr[\xi_n(\tau_r-\tau_h)])$, where $\tau_r$ is the remaining point,
and $\tau_h$ is a point in the hyperplane.
If the two signs are different (i.e. $\rho$ and the remaining element
are ``on different sides''),
then $\rho$ does not belong to $\conv \{\tau_{1}^{(i)},
..., \tau_{L}^{(i)}\}$. This test is performed $L$ times for all 
possible choices of $L-1$ elements from $c_i$. If
 $\rho$ is in each case on the same side as the
remaining point, 
then $\rho$ is in $\conv \{\tau_{1}^{(i)},...\tau_{L}^{(i)}\}$,
and therefore separable. 
In this case the algorithm terminates.
 Otherwise one continues with  the next step, i.e. $i \mapsto
i+1$.

Combining the two algorithms ${\cal A}_1$ and ${\cal A}_2$ 
by running them parallel (in an iterative way)
is already a big improvement
over ${\cal A}_1$, since the combined algorithm terminates
after a finite time for all input states that are entangled and all
input states that are in the interior of the separable states. 
This leaves a set 
at the border between  separable and 
entangled states, where the combined algorithm  cannot be trusted to
terminate after a finite time. 
This problem can be solved as follows.

A state $\rho$ on the border 
between separable and entangled states
has the
property that
for all $0<\eta<1$ the 
operator $\rho_e=(1+\eta)\rho-\eta \eins$ does not belong to the separable
states~\cite{benson}.
Due to convexity the operator
$\rho_s=(1-\eta)\rho+\eta \eins$ is separable.  
If for all $\eta>0$ the operator
$\rho_e$ is non-positive, then $\rho$ is not of
full rank --
a case that we already studied above.
Thus, the only possibility for a state to be
on the border between  separable 
and  entangled states is that
  $\exists \, \eta_0>0$ s.t.
 $\forall \, \eta <\eta_0$ the state $\rho_e$ is positive.
Note that until today
 there is no algorithm known for the decision whether a state
is on the border between separable and entangled states:
If 
 this border
would be known completely
 the separability problem 
would be solved.

The above property can be used for closing the
termination-gap  in 
the combined algorithm described above, 
by extending it in the following way to the final algorithm 
${\cal A}$:
Take some small, but fixed $\eta>0$, such that $(1+\eta)\rho- \eta
\eins$ is a positive operator. Then set two flags $f_1,f_2$ to FALSE. 
These are global flags and are not changed at any step of the algorithm,
unless  mentioned explicitly.
In the $i$-th step of the algorithm:\newline
1. do the $i$-th step of ${\cal A}_1$ for $\rho$,\newline
2. do the $i$-th step of ${\cal A}_2$ for $\rho$,\newline
3. do the $i$-th step of ${\cal A}_1$ for the state $(1+\eta)\rho-
\eta \eins$,\newline
4. do the $i$-th step of ${\cal A}_2$ for the state 
$(1-\eta)\rho+\eta \eins$.\newline
If ${\cal A}_1$ detects $(1+\eta)\rho-\eta \eins$ in 3., set $f_1$ to TRUE 
(from this point on it will stay TRUE).
If ${\cal A}_2$ detects $(1-\eta)\rho+\eta \eins$ in 4., set $f_2$ to TRUE 
(from this point on it will stay TRUE).

The termination criteria for ${\cal A}$ are  given as:\newline
a) If the ${\cal A}_1$ test  detects $\rho$ in 1., then 
$\rho$ is entangled and ${\cal A}$ terminates.\newline
b) If the ${\cal A}_2$ test detects $\rho$ in 2.,
then $\rho$  is separable
and ${\cal A}$ terminates.\newline
c) If both $f_1$ and $f_2$ are TRUE, then the state is in the 
$\eta$-surrounding of 
the border between separable and entangled states,
 and ${\cal A}$ terminates with this
information.\newline
Otherwise do the step $i \mapsto i+1$.

This algorithm  terminates after a finite time for {\em any}
initial  state. 
Here, the 
  outcome c)
 does not 
give any information about the state
being separable or entangled, but just the
knowledge that the state is ``close'' to the border.
However, 
we point out that the surrounding of the border which leads to an
inconclusive outcome can, in principle, be made arbitrarily small.

In summary, we have presented an algorithm for the separability
problem, which complements the algorithm of Doherty,  Parrilo, 
and  Spedalieri. Their algorithm  detects an entangled state
after a finite number of steps, but does not terminate for
separable states. Our algorithm, on the other hand, detects
a separable state after a finite number of steps, but does
not terminate for an entangled state. The connection of the two
algorithms terminates for {\em all} input states. In the case of the
initial state being close to the border between  separable and
 entangled states, our algorithm terminates with an inconclusive
output.

We acknowledge discussions with M. Lewenstein and 
support from DFG via 
Schwerpunkt 1078 ``Quanten-Informationsverarbeitung'',
SFB 407 `` Quantenlimitierte Messprozesse mit Atomen, Molek\"ulen 
und Photonen''
and 
European Graduate College 665 ``Interference and Quantum Applications".

\end{document}